# Exploiting Locality in Searching the Web

Joel Young　　Thomas Dean
Computer Science Department
Brown University, Providence, RI 02906

## Abstract

Published experiments on spidering the Web suggest that, given training data in the form of a (relatively small) subgraph of the Web containing a subset of a selected class of target pages, it is possible to conduct a directed search and find additional target pages significantly faster (with fewer page retrievals) than by performing a blind or uninformed random or systematic search, e.g., breadth-first search. If true, this claim motivates a number of practical applications. Unfortunately, these experiments were carried out in specialized domains or under conditions that are difficult to replicate. We present and apply an experimental framework designed to reexamine and resolve the basic claims of the earlier work, so that the supporting experiments can be replicated and built upon. We provide high-performance tools for building experimental spiders, make use of the ground truth and static nature of the WT10g TREC Web corpus, and rely on simple well understand machine learning techniques to conduct our experiments. In this paper, we describe the basic framework, motivate the experimental design, and report on our findings supporting and qualifying the conclusions of the earlier research.

## 1 Introduction

The Web is large, rapidly growing and full of information people need. While commercial search engines, or trawlers, work for many information-gathering purposes, they have limitations. Without belaboring those limitations, we assume that there is value to be had in finding pages that haven't yet been indexed, avoiding hidden biases in the indexing and ranking strategies used by trawlers, finding pages that can't be characterized in terms of a list of keywords, and searching portions of the Web (private intranets, robot exclusions) that are inaccessible to trawlers.

In recent years there have been a number of proposals for customizable search engines, called spiders, that are computationally lean (they rely on one or two workstations rather than a roomful of powerful servers), narrow in focus (they can answer highly specialized queries), informationally thrifty (they rely on small amounts of local information to guide search) and adaptive (they learn to search based on a sampling of the Web rather than exhaustive search) (Chakrabarti, van den Berg, and Dom, 1999; Chau, Zeng, and Chen, 2001; Diligenti et al., 2000; Najork and Wiener, 2001; Rennie and McCallum, 1999). The papers describing these spiders typically report some degree of success searching specialized domains, answering ad hoc queries, or rely on methods that are difficult to replicate.

In this paper, we[1] describe an experimental framework for investigating the questions raised by the earlier work, report on our evidence supporting the earlier results, and provide a somewhat more nuanced view of the promise for this approach to searching the Web. Our framework includes a powerful set of tools for building and deploying spiders; these tools were designed with the experiments in this paper in mind but already have proved useful for other experiments searching the Web and analyzing its properties (Pandurangan, Raghavan, and Upfal, 2002). To avoid the generality and replication problems of the earlier experimental work, we rely on the ground truth and stability provided by the WT10g Web Corpus (CSIRO, 2001). Our tools use proxy layers to simplify switching between various corpora and the wild Web.

We begin in Section 2 by stating our primary hypoth-esis concerning the possibility of efficient local search. In Section 3, we examine properties of the Web that might facilitate such search. In Section 4 we describe our experiments and analyze the results. Section 5 describes the software architecture developed to support our experiments and now available to the research community. Finally, in Section 6 we review the experimental evidence and offer some additional insights.

---

[1] The views expressed in this article are those of the authors and do not reflect the official policy or position of the United States Air Force, Department of Defense, or the U.S. Government

## 2  Efficient spidering hypothesis

The Web can be viewed as a directed graph in which pages are the nodes and hyperlinks are the edges. One can readily imagine a robot traversing this graph by selecting at each node which out-going edge to take next. Unlike embodied robots, a web robot (appropriately called a *spider*) can teleport to any location whose URL it knows, at a cost no more than that incurred in traversing a single edge. A spider taking advantage of this property has at each step the choice of exploring any node in the graph it knows how to get to. Those nodes that are known, but unvisited, are called *fringe nodes*. The set of all such nodes is called the *fringe*. In the remainder of this paper when we refer to a "page" think "node in a web graph" and when we refer to a "link" or "URL" think "directed edge in a web graph."

A *target* page is simply a member of a subset of the pages in the web graph. We're interested subsets representing reasonably coherent topics or themes — for example, pages about William James' philosophical views on determinism and free will. We seek an answer to the following question:

> Given training data in the form of a (relatively small) subgraph of the Web containing a subset of the target pages, is it possible to conduct a directed search and find additional target pages significantly faster (with fewer page retrievals) than by performing a blind or uninformed random or systematic search, such as depth-first or breadth-first search?

This question, which we call the "the efficient spidering hypothesis," addresses the feasibility of the following concrete task. Suppose you're given a set of representative pages, $p_1, ..., p_n$, from an unspecified (but thematically coherent) subset. By submitting these pages to a spider as a "query by example," you initiate the following three-phase search process: In the initial *exploratory phase*, the spider examines additional pages in the vicinity of $p_1, ..., p_n$, possibly using the ability to find "back links" provided by exhaustive search engines. In the *learning phase*, the spider uses those portions of the web graph found in the exploratory phase to construct a function for ranking nodes in the fringe to guide a subsequent directed search. In the final *performance phase*, the spider conducts a search guided by this ranking function, possibly using various strategies for random restarts or redirection to avoid "over grazing" portions of the Web or finding the same pages over and over again.[2] After visiting a predetermined number of pages, the spider displays the pages ordered according to their similarity to $p_1, ..., p_n$ or their distance from pages "like" $p_1, ..., p_n$. While relevant to building spiders, we don't directly address the problem of target identification in this papers.

In the simplest experiments designed to resolve this question, different spiders are evaluated in terms of the total targets found in a fixed number of actions and then compared to non-informed search methods. Other experiments involve using more revealing performance measures and controlling for the number of representative pages, the size of the subgraph explored in the exploratory phase, the method used in the learning phase, and the search strategy used in the performance phase. Before we describe our experiments, let's consider some characteristics of the Web that might support efficient spidering.

## 3  Properties of the Web

The Web — or at least the portion that we're interested in — is connected and has relatively low *average* diameter.[3] However, while there may be a short path from where you are to where you want to be, finding such a path may be difficult given the relatively high average out-degree of nodes in the connected part of the Web. Fortunately, the Web seems to be consistently structured in the sense that a coherent set of web-page authoring and linking strategies are pervasive. Our confidence in the feasibility of efficient spidering is based in part on our assumption that the rules governing the creation of content (including the selection, arrangement and frequency of appearance of terms) and the citing (or linking) of pages are statistically similar throughout the relevant fragments of the Web.

The search problem faced by a spider involves a series of dependent decisions. At each decision point in its performance phase, the spider has information local to the parts of the Web it has seen so far. The spider has seen several target-related portions of the Web during its exploratory phase; hence, if we are correct in our assumption regarding the statistical regularity of page generation and linking, there should be clues available locally that, combined with the spider's analysis of similar portions of the Web, should provide a more global perspective. The efficient spidering hypothesis is concerned with whether or not a spider can achieve such a global perspective by exploring regions of the web graph containing examples of target pages.

---

[2] While the problem of over grazing is potentially serious and practical spiders will likely adopt strategies to avoid it, the experiments in this paper deal with spiders guided entirely by ranking the pages in their fringe.

[3] Kumar et al. report that, for the approximately 25% of pairs of pages for which there exists a connecting path, the average path length following in-links, out-links or both is, respectively, 16.12, 16.18 and 6.83. (Soboroff, 2002) provide evidence that WT10g exhibits similar properties.

We don't assume that there will necessarily be a short path connecting two pages that address the same topic — such a shortcut may eventually be created, but we're often interested in finding isolated pages prior to these pages being identified by their interested communities. We do assume, however, that if multiple clusters of pages are created around a topic or set of topics, then the authors of these pages will arrange the relevant subgraphs in a semantically similar fashion, for example, on average, pages with content of type $A$ will be linked to pages with content of type $B$ in both clusters. We don't pretend to know exactly how to characterize these "navigationally-relevant" types of content; however, in order to answer our question concerning efficient spidering in the affirmative, we are relying on relatively standard supervised learning algorithms to reveal the underlying navigational structure. In order to outperform other search methods, the ranking function guiding a smart spider has to exploit navigational cues in the content of pages to compute a gradient pointing to nearby target pages.

## 4 Experiments

Earlier work (much of it conveniently summarized in (Chakrabarti, 2002)) provides some experimental evidence supporting the efficient spidering hypothesis, but the data is problematic: targets are often contrived, the experimental situation is typically inadequately described, and the opportunities for replication are few and unsatisfying. In order to provide solid evidence confirming or disconfirming the hypothesis, we need to

- fix the experimental environment — define a subgraph of the web graph in which to run experiments — without the stability of a fixed environment it is difficult to compare algorithms and impossible for other researchers to validate claims, replicate and extend experiments,

- establish ground truth — identify a set of target classes and then find all pages in that class within the fixed environment — without this foundation, measuring performance must rely on ad hoc target identification criteria and incomplete target information, and

- quantify the experimental environment — provide tools to determine properties of relevant subgraphs such as their connectivity and intersection — without these tools it is impossible to interpret the results of experiments.

All of these requirements are either directly solved or facilitated by using the WT10g TREC Web Corpus.

Another problem with interpreting results in the literature concerns variation in performance measures. In our work, we use two of the most common measures (time to first hit — the number of URLs fetched prior to first target — and the total number of hits given a fixed number of fetches) plus one additional measure (discounted cumulative reward — sum of the rewards $\sum_t r_t \gamma^t$ where $t$ is the time measured in fetches, $r_t$ is the reward at time $t$ (1 for a hit and 0 otherwise), and $\gamma$ is the discount rate) which we argue in (Young, 2003) constitutes an appropriate gold standard.

Relative to the efficient spidering hypothesis, we concentrate on three non-informed or systematic methods: random walk (Random selects randomly from the set of all fringe nodes, depth-first search (DFS) selects randomly from the set of pages in the fringe at the next level/depth, exhausting that level before considering pages at the current level, and breadth-first search (BFS) selects randomly from the set of pages at a given level, exhausting each level before considering pages at the next level.

We separate the experimental procedure into four stages: setup, training, testing, and evaluation. In the setup stage, we select a target class, a set of targets from this class to be used for training, a set of targets for testing, and a set of starting pages to sample from. In some experiments, we sample only from those pages known to be $k$ steps from the nearest target. There are also concerns about overlap between the neighborhoods of the training and testing targets that we'll address momentarily. In this paper, we focus on TREC target class #544 against the WT10g corpus which is characterized as "documents that describe the roles estrogen plays in the human body." #544 consists of 324 targets and we use 13 for training and the remainder for testing.

Training proceeds by starting with a (training) set of target pages and traversing backlinks to find a collection of additional pages in the vicinity of the target pages. Backlinks are found using WT10g or, in the case of searching the wild web, either Google's advanced search services or by using the spider to collect pages through forward links and reconstructing a por-tion of the web graph[4]. A set of pairs of the form $(p, z)$ (a supervised training example) is found where $p$ is a page and $z$ is either the length of the shortest path to a target (termed *depth* for short) in the training set or the accumulated discounted reward calculated according to the obvious Markov model for the underlying planning problem (Young, 2003). While obvious in our estimation, this model doesn't correspond to the classic formulation for stationary finite-state Markov decision processes; in the obvious model where states correspond to pages, the rewards for spidering are non-stationary (you don't receive a reward for finding a target the second time). There are alternative models, e.g., (Rennie and McCallum, 1999), but they rely on convoluted state and actions models. This characteristic of the problem poses complications for solution methods, such as reinforcement learning, that depend on stationarity.

|  | Machine | Median $\Delta$ Discount | Mean $\Delta$ Discount |
|---|---|---|---|
| Gold | Random | 94 | 292 |
| Depth | Breadth | 139 | 278 |
|  | Depth | 98.4 | 346 |
| Gold | Random | 94.1 | 292 |
| Discount | Breadth | 139 | 278 |
|  | Depth | 98.5 | 346 |

Table 1: 95 start pages from 4 away to depth 5. *gold* standard vs. non-informed. Discounts are $log_{10}$ with discount $\gamma = 0.5$

Before we invest in training spiders using either the depth or discounted reward heuristics, we should test if the heuristic works for spidering. One can imagine a scenario in which these heuristics don't efficiently guide a spider as our calculations of the depth or reward don't take into account the nonstationary nature of the process, that is, we calculate the reward or depth of a training page as if it was the first page visited. In an actual spidering run, we may think a new page has high reward when in fact we have already visited the targets reachable from it. This does not turn out to be a problem however. In Table 1 we see the results from a set of runs comparing the performance of spiders using the heuristics vs. randomized runs. We perform 20 randomized runs for each of the un-informed search strategies. According to the two-sample paired (Wilcoxon) signed rank test of the null hypothesis that the medians of the paired differences in performance are zero, the probability of observing the median differences we observe is 0. We therefore accept the alternate hypothesis that the depth and re-ward heuristics strongly outperform the un-informed searches. These are our *Gold Standards*. Although we show only one data point, we observed the same dominance in our gold standards in all other trials.

We experimented with a number of learning methods including simple regression, decision trees and support vector machines. The results reported here make use of one of the simplest, most generic methods for learning: linear-kernel Support Vector Machines (SVMs). We also used very basic methods from information retrieval to encode pages as word frequency vectors. These methods include standard methods for stemming and stopword pruning. We also use information gain to reduce dictionary size (from approximately 600,000 down to 10,000 terms) and thereby obtain a reasonable cap on the dimensionality of the word frequency vectors. The methods that we employed are common within the IR community and the gory details along with the parameter settings for the SVM libraries (Joachims, 1999; Chang and Lin, 2002; McCallum, 2002) we used are provided in (Young, 2003).

In the testing stage, we randomly select a set of starting pages from an appropriate subset of the set of all pages, run each method (random, BFS, DFS, SVM) from each starting page for a fixed number of page retrievals. For diagnostic purposes, in some experiments we limit the depth of all methods in order to emphasize the strengths and weaknesses of the various methods.

During the evaluation stage, we measure the average pairwise differences between methods and then use the Wilcoxon test to test against the null hypothesis that the population median of the pairwise differences is 0.

One aspect of evaluation that is typically missing in other reported experiments, is consideration for the problem of possible overlap between training and testing data. Dealing with this problem is particularly complicated in the spidering problem. The *light-cone* of a target page refers to the set of pages linked to the target page by (shortest) paths of length 1, 2, 3, and so on. The following imagery helps to illustrate the overlap problem (a page $n$ steps away from the target refers to a page such that there is a (directed) path traversing hyperlinks of length $n$ from the page to the target):

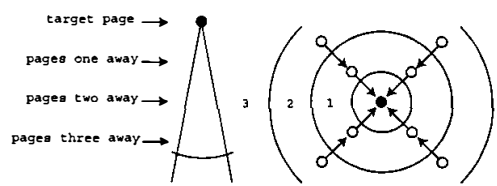

---
[4] The set of pages reachable in depth 6 from the the set of positive targets for a TREC task overlaps by up to 70% with the depth 4 backlink page set. This suggests that automated access to backlinks is not required.

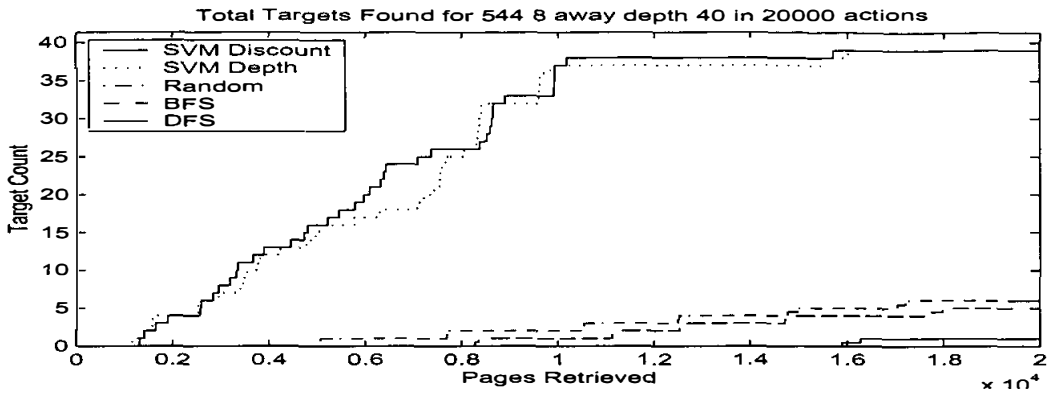

Figure 1: 101 starting pages at 8 away to depth 40 for 20000 page retrievals. Median targets found

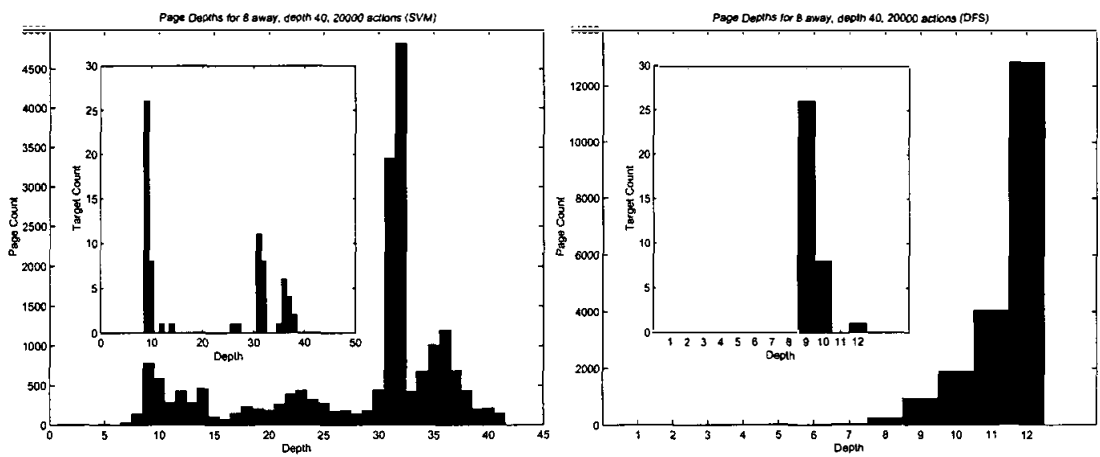

Figure 2: Distribution of pages and targets (insets) for one SVM and matching BFS spider started from the same page from Figure 1

We can now ask how the light-cones of depth $n$ emanating from targets in the training set overlap with those for targets in the testing set. This and similar statistics provide insight into how the test and training data are separated as well as insight into the TREC target classes themselves. For the #544 (estrogen-related) target class, there is remarkably little overlap out to a depth of five or six for the randomly selected training and testing sets used in our experiments. But this isn't true for all target classes.

We're also interested in how those parts of the graph traversed by the spider during testing overlap with that portion of the graph used for training. In this case, we are interested in an inverted light-cone with edges leading away from a starting page

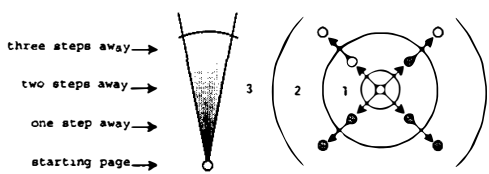

where only those pages at depth $n$ are actually included in the cone, and the relevant potential for overlap can be depicted thus

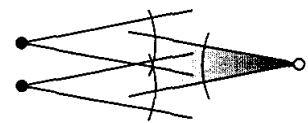

In our experiments, we have taken pains to avoid such overlap where possible and quantify it to the extent that it does occur. For the experiments involving the #544 target class, there is never more than a 10% overlap, and, in the overwhelming majority of cases, there is zero or negligible overlap. A combination of the WT10g corpus and the instrumentation available in our spider software makes it simple to gather the necessary data to compute such statistics.

Given this lengthy prologue, here is our first figure (Figure 1) showing results of 101 runs starting from pages known to be a distance of 8 from the nearest target page and allowed to make 20000 fetches. The top two lines are the median total targets found by the SVM trained spiders. Then the next best performing line is the random search followed by breadth first search. Depth first comes in way last. Table 2 gives the amount the trained spiders beat the non-informed strategies and the Wilcoxon test statistic at the end of 20,000 retrievals. The difference in performance is significant.

|  | Non-Informed | Median Δ Targets | Wilcoxon Signed Rank $p$ |
|---|---|---|---|
| Depth SVM | Random | 33 | 6.2e-35 |
|  | Breadth | 34 | 6.0e-35 |
|  | Depth | 38 | 3.0e-35 |
| Discount SVM | Random | 34 | 6.3e-35 |
|  | Breadth | 35 | 5.9e-35 |
|  | Depth | 39 | 3.6e-35 |

Table 2: 101 starting pages at 8 away with maximum search depth of 40 after 20000 retrievals. Depth and discounted reward trained SVMs vs. Random, BFS, and DFS

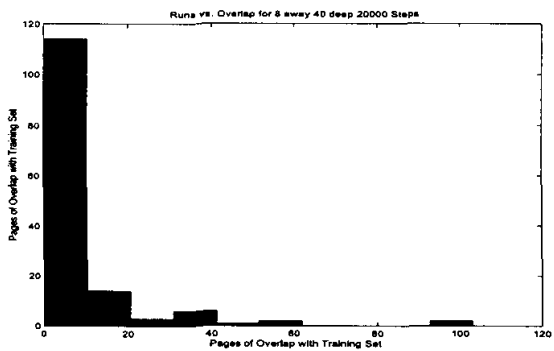

Figure 3: Distribution of overlap between the test spider runs from Figure 1 and the training light-cone.

What is the SVM spider doing to beat the other strategies? These histograms in Figure 2 breakdown the depths at which two spiders, starting from the same start page 8 away from the closest target and running for 20000 steps, discover pages and find targets. Examine the top two histograms. Notice the three humps where the SVM finds targets at 9, 30, and 36. notice the focus on exploring pages at those depths. The bottom two histograms show the pages explored by a DFS spider. Notice that it is finding the same targets at depths 9 and 10 but that the SVM is able to keep on going beyond that first pile of targets to find two more sets.

The explored regions for all of these runs from Figure 1 had less than 0.5% overlap with the training light-cone. Figure 3 shows the small amount of overlap. Note that the starting pages were not filtered to gain this property. Just as the low amount of overlap between test and training These performance curves are typical. Further examples and discussion are available in (Young, 2003).

These results portray the power of the SVM spiders on typical spider tasks where the depth is not constrained. However, with this type of experiment, the domain is extremely large (in the hundreds of thousands of pages) and is intractable to search enough times to build statistics. In the following results we perform exhaustive searches of much smaller domains. We start the spiders at a fixed distance to the nearest target and constrain the spider to searching ahead no greater than a small distance. This has the side affect of constraining the smart spiders from following a "scent" until a shorter path is discovered.

| SVM | Non Informed | Median Δ (Mean) | Wilcoxon Signed Rank $p$ |
|---|---|---|---|
| Depth | Random | 16.6( 148) | 4.1e-23 |
|  | BFS | 40.4( 134) | 8.5e-90 |
|  | DFS | 19.2( 202) | 1.5e-30 |
| Discount | Random | 17.8( 148) | 4.5e-25 |
|  | BFS | 40.2( 134) | 5.4e-96 |
|  | DFS | 21.1( 202) | 1.0e-31 |
| Depth | Discount | 0 (-0.011) | 5.0e-6 |

Table 3: 95 starting pages at 4 away to depth 5. $\log_{10}$ of 0.5 Discounted Reward

| SVM | Non Informed | Median Δ (Mean) | Wilcoxon Signed Rank $p$ |
|---|---|---|---|
| Depth | Random | -55( -493) | 4.6e-23 |
|  | BFS | -134( -445) | 1.0e-89 |
|  | DFS | -134( -445) | 1.0e-89 |
| Discount | Random | -59( -492) | 4.9e-25 |
|  | BFS | -133( -444) | 5.5e-96 |
|  | DFS | -133( -444) | 5.5e-96 |
| Depth | Discount | 0 (-0.01) | 0 |

Table 4: 95 starting pages at 4 away to depth 5. Actions to First Target

Table 3 shows the performance of our different spidering strategies as measured by median difference in $\log_{10}$ of the 0.5 discounted reward with the mean of same in parenthesis. We read the table as follows: The median difference in performance between the depth trained SVM spider and Random spider is 55 with significance with a mean difference of 493. We can see from this table that the SVM trained spiders outperformed, with significance, all of the non-informed search strategies. We can also see that the depth trained and discounted reward trained SVM spiders performed equivalently in median, but without

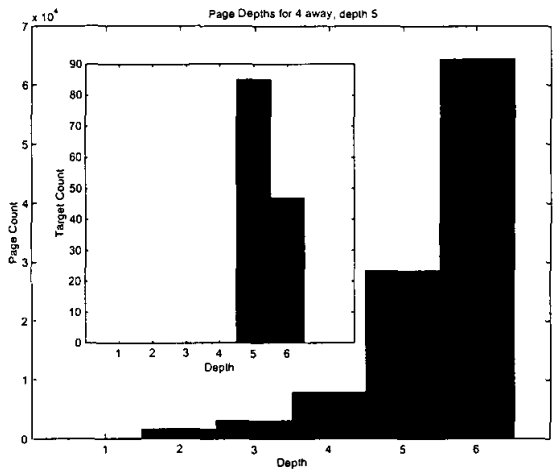

Figure 4: Distribution of Pages and Targets (inset) for 95 starting pages at 4 away to depth 5

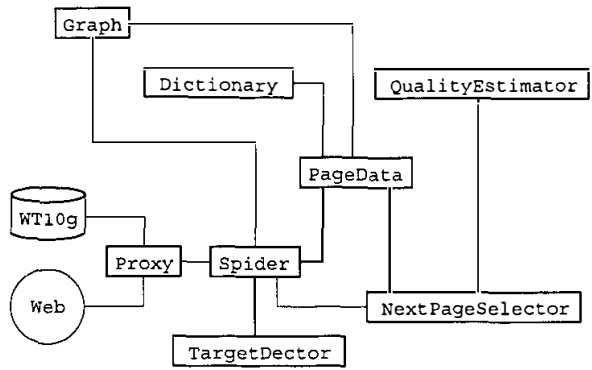

Figure 5: Components of the spider architecture

significance—examination of the data reveals it is completely one-sided in favor of the discounted spider. Table 4 shows number of actions to first target instead. For example, the discounted reward trained SVM spider finds, in median, the first target page 59 steps sooner than a random spider with significance. Negative value for the difference indicates SVM machine performed better than the non-informed machine.

To get a feel for the space we are searching in examine Figure 4. The figure shows how the pages and targets are distributed. As expected, we see a geometric growth in page count at each depth with the expected linear growth found in web graphs. We can also see that most of the targets to be found are clustered at depth 4.

|  | Non | Median $\Delta$ | Wilcoxon |
|---|---|---|---|
| SVM | Informed | (Mean) | Signed Rank $p$ |
| Depth | Random | 6( 14.3) | 1.05e-198 |
|  | BFS | 1.81( 7.71) | 1.73e-118 |
|  | DFS | 12.9( 34.1) | 7.99e-283 |
| Discount | Random | 6.07( 14.4) | 4.69e-203 |
|  | BFS | 1.93( 7.86) | 1.03e-118 |
|  | DFS | 13( 34.2) | 7.82e-288 |
| Depth | Discount | 0 (-0.149) | 0.0376 |

Table 5: 256 Target pages to depth 5. $log_{10}$ of 0.5 Discounted Reward

Giving another snapshot of performance, Table 5 shows our spiders starting right at target pages. This tests the ability of the spiders to *exploit* target rich areas rather than *exploring* to find new target clusters.

In the above experiments, we showed the trained spiders ability to search in constrained environments better than the non-informed techniques.

## 5 Spider software architecture

Figure 5 shows the basic software components for the spider toolkit. The Spider module provides the main search loop for the spider. Various layered components are provided allowing the user to design a spider with the particular properties needed. It also manages all of the resources used. The Spider module communicates with an http proxy. The Proxy module hides the issues with accessing the real Web or the WT10g corpus or other homegrown corpora. The Spider module stores the information about the Web it discovers in the Graph and PageData modules. The Graph module leverages the BOOST Graph Library (Siek, Lee, and Lumsdaine, 2003) to allow efficient representation and ready access to a broad collection of graph algorithms. The Spider module derives its search strategy from the NextPageSelector. The NextPageSelector manages the fringe and selects pages from the fringe according to the QualityEstimator. Different pluggable QualityEstimator modules provide DFS, BFS, Random, SVM, .... The QualityEstimator abstracts the spider from the machine learning technique used. By writing a small module, many third party machine learning modules can be attached. Modules have been written for such systems as C4.5, BOW, libsvm, and svmlight. The TargetDetector isolates the task of identifying target pages. Modules have been written using a thresholded inner-product similarity as well as an oracle look-up into a known target list.

The modular architecture lets us readily adapt a spider for different environments. We can plug-in a new parser for backlink detection from Google or Alta Vista, we can plug-in different filters to limit the spider's environment, e.g., avoiding binary pages or staying in the .gov domain. With a command line switch, we can switch from the live-web to the WT10g corpus.

## 6 Discussion

Note that despite the analysis in Section 4 it may still be the case that over some portions of the web graph or for some topics or sets of target pages there are no discernible local signals to guide search. Such signals may take time to manifest; informative signals may emerge at different times as topics and communities mature. In such weak-signal backwater areas, it may prove necessary to "amplify" the signal by broadening the definition of "local" to refer to all pages within a small diameter. It may also be the case that in some areas the signals are simply nonexistent.

What we know now that we didn't before is, at least for some topics, the signal is stronger and more pervasive than the earlier research gave us any reason to suspect. We have been able to scrutinize the distributional and connectivity properties of some of the target classes in the WT10g corpus, but more work is needed to better understand how these properties affect the behavior of spiders guided by local search methods. Also, it's worth pointing out that our results don't depend on any particularly sophisticated learning methods and, hence, we might hope for better performance using more state-of-the-art methods.

And, finally, the algorithms and data for our experiments will be made publicly available. A link to an archive file containing all the code and basic instructions for building the various tools described in this paper will be ready by late spring or early summer. In the meantime, the first author will make the code available to interested researchers. We invite others to replicate and extend our experiments and, perhaps bolstered by our repeatable and controlled experiments, perform additional experiments on other corpora and the wild Web.